\documentclass[aps,preprint,showpacs,superscriptaddress,groupedaddress]{revtex4} 

\usepackage[dvips]{graphicx}
\usepackage{dcolumn}   
\usepackage{bm}        
\usepackage{amssymb}   
\usepackage[utf8]{inputenc}
\usepackage{mathrsfs}
\usepackage{amsmath}
\usepackage[english]{babel}

\usepackage[colorlinks]{hyperref}

\hypersetup{colorlinks=true,citecolor=blue,linkcolor=blue,urlcolor=blue}
\usepackage[usenames]{color}
\usepackage{cancel}

\begin{document}

\preprint{\small Chaos, Solitons and Fractals \textbf{165} (2022) 112805   
\ [\href{https://doi.org/10.1016/j.chaos.2022.112805}{DOI: 10.1016/j.chaos.2022.112805}]
}


\title{Kinks in higher-order polynomial models}

\author{Petr A. Blinov}
\email{blinov.pa@phystech.edu}
\affiliation{Moscow Institute of Physics and Technology,
Dolgoprudny, Moscow Region 141700, Russia}

\author{Tatiana V. Gani}
\email{gani.t@bk.ru}
\affiliation{Faculty of Physics,
M.~V.~Lomonosov Moscow State University, Moscow 119991, Russia
}

\author{Alexander A. Malnev}
\email{malnev.aa@phystech.edu}
\affiliation{Moscow Institute of Physics and Technology,
Dolgoprudny, Moscow Region 141700, Russia}

\author{Vakhid A. Gani}
\email{vagani@mephi.ru}
\affiliation{Department of Mathematics, National Research Nuclear University MEPhI (Moscow Engineering Physics Institute),
Moscow 115409, Russia}
\affiliation{Kurchatov Complex for Theoretical and Experimental Physics\\
of National Research Centre ``Kurchatov Institute'', Moscow, Russia}

\author{Vladimir B. Sherstyukov}
\email{shervb73@gmail.com}
\affiliation{Faculty of Mechanics and Mathematics, M.~V.~Lomonosov Moscow State University, Moscow Center for Fundamental and Applied Mathematics, Moscow 119991, Russia
}


\begin{abstract}
We consider a family of field-theoretic models with a real scalar field in (1+1)-dimensional space-time. The field dynamics in each model is determined by a polynomial potential with two degenerate minima. We obtain exact general formulas for kink solutions with power-law asymptotic behavior. We also write out formulas for the asymptotics of all found kinks. In addition, we analyze some other properties of the obtained kinks: stability potentials, zero modes, positions of the centers of mass.
\end{abstract}



\maketitle


\section{Introduction}
\label{sec:Introduction}

Kink solutions are one of the types of topological solitons. They are found in a large number of $(1+1)$-dimensional field-theoretic models, see, e.g., books and reviews \cite{Manton.book.2004,Shnir.book.2018,Rajaraman.book.1982}. Over the last few years, there has been a growing interest in kinks with power-law asymptotics, which arise in models with both polynomial and non-polynomial potentials.

Models with polynomial potentials and their kink solutions are widely applied in physics. For example, the famous $\varphi^4$ potential models spontaneous symmetry breaking. Such a potential arises in the Ginzburg--Landau theory of superconductivity \cite{Ginzburg.ZhETF.1950}. Deformations in the form of the $\varphi^4$ kinks were found in buckled graphene \cite{Yamaletdinov.PRB.2017}, and some further non-trivial phenomena were observed in the interactions of kinks in buckled graphene nanoribbons with radiation \cite{Yamaletdinov.Carbon.2019}. Models with sixth degree polynomial potential are used to describe structures in tetragonal ferroelectric perovskites \cite{Cao.PRB.1991}.

Higher-order polynomial potentials and corresponding kink solutions with power-law asymptotics are also in demand in many physical contexts. In particular, such theories are used to describe sequences of phase transitions in some materials, see, e.g., Ref.~\cite{Khare.PRE.2014} or chapter 12 in the book \cite{Kevrekidis.book.2019} (the text of this chapter can also be found in arXiv e-print \cite{Kevrekidis.chapter.2018}) and relevant references therein. Besides that, there are cosmological applications of higher-order polynomial potentials in situations where it is required to simulate the appearance of additional vacua in the theory \cite{Greenwood.PRD.2009}.

Nonlinear physicists are always striving to find exact solutions. However, quite often this is not possible. Or it is possible, but the result is huge formulas obtained using computer algebra systems and not amenable to analysis and comprehension. The latter option also seems not very suitable.

As for field-theoretic models with polynomial potentials, the situation here does not look entirely pessimistic. In addition to the well-known kink in the $\varphi^4$ theory \cite{Kevrekidis.book.2019}, which has been actively studied since the mid-1970s, we mention some later results.
\begin{itemize}
    \item In 1979 M.A.~Lohe published a well-known paper \cite{Lohe.PRD.1979} where the kinks of the $\varphi^6$ model with three vacua were written down. In this paper, for one of the variants of the $\varphi^8$ model with four vacua, an implicit expression for kinks $x=x_{\scriptsize\mbox{K}}^{}(\varphi)$ is given.
    \item An interesting approach based on the so-called {\it deformation procedure} \cite{Bazeia.PRD.2002} was used in Ref.~\cite{Bazeia.PRD.2006}. In this paper, kink solutions are obtained in a family of models with polynomial potentials of a very special form. The $\varphi^4$ was used as the initial model, which was then subjected to ingenious deformations.
    \item In 2014 in paper \cite{Khare.PRE.2014}, implicit solutions were obtained for various variants of the $\varphi^8$, $\varphi^{10}$ and $\varphi^{12}$ models.
    \item In Ref.~\cite{Gani.PRD.2020} it was shown that finding kink solutions to the $\varphi^8$ model with the potential $V(\varphi) = \displaystyle\frac{1}{2}\left(\varphi^2-a^2\right)^2\left(\varphi^2-b^2\right)^2$ reduces to solving algebraic equations of a certain form. In some cases, these equations can be solved and the kink can be obtained explicitly as $\varphi=\varphi_{\scriptsize\mbox{K}}^{}(x)$.
    \item A family of symmetric polynomial potentials with two and three degenerate minima is considered in Ref.~\cite{Khare.JPA.2019}. Implicit expressions for kinks are found and their asymptotics are obtained.
\end{itemize}

We also mention several important results obtained in the last few years and related to kinks with power-law asymptotics.
\begin{enumerate}
    \item The conditions for the potential are formulated, which lead to the appearance of kinks with one or two power-law tails \cite{Christov.PRD.2019}.
    \item It is shown that the presence of power-law tails leads to long-range interaction in kink-kink and kink-antikink systems \cite{Christov.PRD.2019,Christov.PRL.2019,Radomskiy.JPCS.2017}. Various approaches have been successfully applied to calculate the interaction forces \cite{Christov.PRL.2019,Manton.JPA.2019,dOrnellas.JPC.2020,Campos.PLB.2021}. For example, in Ref.~\cite{Christov.PRL.2019} for kinks with one power-law and one exponential tail, it is shown that in the $\varphi^{2n+4}$ models with potentials $V(\varphi)=(1-\varphi^2)^2\varphi^{2n}$ for $n>1$, the kink and antikink faced to each other by power-law tails, attract, while the kink and kink repel. For the attraction/repulsion force of two kinks, a power-law dependence of the force on distance was found.
    \item It is shown that for the numerical simulation of kink-(anti)kink scattering processes in the case of power-law tails, the usual (traditionally used in the case of kinks with exponential asymptotics) formulation of the initial conditions is not applicable \cite{Christov.PRD.2019}. Improved initial conditions were developed and successfully applied using an additional minimization procedure, which made it possible to exclude artifact radiation from the tail overlap region. This radiation strongly affects the results of numerical experiments, in particular, leads to the illusion of repulsion between the kink and antikink \cite{Christov.PRD.2019,Belendryasova.CNSNS.2019}.
    \item In collisions of a kink and an antikink with power-law tails, resonant phenomena were discovered --- {\it escape windows} forming a quasi-fractal structure \cite{Christov.CNSNS.2021}. This is an interesting fact, since the excitation spectrum of a kink with at least one power-law tail contains no vibrational modes that could play the role of an energy accumulator. Several hypotheses have been put forward \cite{Belendryasova.CNSNS.2019,Gani.JPCS.2020.no-go}, however, apparently, no final solution has been found yet.
    \item The transformation properties of the kink asymptotics with respect to the deformation procedure have been studied \cite{Blinov.AoP.2022,Blinov.JPCS.2020.deform}. A class of deformation functions has been found that transform the power-law asymptotics into power-law, but it was shown that the speed of the field approaching the vacuum value (the power of the coordinate in the asymptotic expression for the field) can change.
\end{enumerate}

This paper fills a gap in the literature regarding topological solitons in models with polynomial potentials. We have obtained exact general formulas for kink solutions in a wide class of models with polynomial potentials with two minima, studied their properties that are important for applications: asymptotics, masses, stability potentials, zero modes. For asymmetric kinks, we have also found the position of the center of mass, which is a non-trivial characteristic. Taking into account the asymmetry of the kink and the location of its center of mass can facilitate the formulation of initial conditions for the numerical simulation of multikink processes, see, e.g., \cite{Moradi.JHEP.2017,Gani.EPJC.2019,Gani.EPJC.2021}. It is noteworthy that in our study we faced with such mathematical structures as Bernstein polynomials and Pascal trapezoids.

The paper is organized as follows. In Section \ref{sec:Models}, we give basic information about field-theoretic models in $1+1$ dimensions and their kink solutions. Section \ref{sec:Solutions} presents our main results obtained for a class of models with polynomial potentials with two degenerate minima. First, in Subsection \ref{sec:Case_A}, the case of kinks with one power-law and one exponential asymptotics is considered, then in Subsection \ref{sec:Case_B}, symmetric kinks with power-law asymptotics are discussed, and finally in Subsection \ref{sec:Case_C} asymmetric kink solutions with both power-law asymptotics are presented. We conclude with a brief discussion of the obtained results and prospects for further research in Section \ref{sec:Conclusion}. Some technical details are included in Appendices \ref{sec:Appendix_A}, \ref{sec:Appendix_B}, \ref{sec:Appendix_C}, and \ref{sec:Appendix_D}.

\section{General information on kink solutions}
\label{sec:Models}

Consider a field-theoretic model with one real scalar field in two-dimensional space-time, which is defined by the Lagrangian
\begin{equation}\label{eq:lagrangian}
    \mathcal{L} = \frac{1}{2}\left(\frac{\partial \varphi}{\partial t}\right)^2 - \frac{1}{2}\left(\frac{\partial \varphi}{\partial x}\right)^2 - V(\varphi),
\end{equation}
where $V(\varphi)$ is a {\it potential} which defines self-interaction of the field $\varphi(x,t)$. The Lagrangian \eqref{eq:lagrangian} leads to the equation of motion
\begin{equation}\label{eq:eqmo}
    \frac{\partial^2\varphi}{\partial t^2} - \frac{\partial^2\varphi}{\partial x^2} + \frac{dV}{d\varphi} = 0.
\end{equation}
In physical applications one is usually interested in non-negative potentials that have two or more degenerate minima $\varphi_1^{{\scriptsize \mbox{(vac)}}}$, $\varphi_2^{{\scriptsize \mbox{(vac)}}}$, ..., with $V(\varphi_1^{{\scriptsize \mbox{(vac)}}})=V(\varphi_2^{{\scriptsize \mbox{(vac)}}})=...=0$. The energy functional for the field $\varphi(x,t)$ is
\begin{equation}\label{eq:energy}
    E[\varphi] = \int\limits_{-\infty}^{\infty}\left[\frac{1}{2} \left( \frac{\partial\varphi}{\partial t} \right)^2 + \frac{1}{2} \left( \frac{\partial\varphi}{\partial x} \right)^2 + V(\varphi)\right]dx.
\end{equation}

In a static case, Eq.~\eqref{eq:eqmo} takes the form
\begin{equation}\label{eq:steqmo}
    \frac{d^2\varphi}{dx^2} = \frac{dV}{d\varphi},
\end{equation}
and the energy looks like
\begin{equation}\label{eq:stenergy}
    E[\varphi] = \int\limits_{-\infty}^{\infty}\left[\frac{1}{2} \left(\frac{d\varphi}{dx} \right)^2 + V(\varphi)\right]dx.
\end{equation}
In what follows, we will be interested in static {\it topological} configurations with finite energy, hence at $x\to\pm\infty$ the field tends to different vacuum values. Moreover, we will consider potentials with two minimum points $-1$ and $+1$, and will be interested in {\it kink solutions} $\varphi_{\scriptsize\mbox{K}}^{}(x)$ for which
\begin{equation}\label{eq:two}
    \lim_{x\to-\infty}\varphi_{\scriptsize\mbox{K}}^{}(x) = -1,
    \quad
    \lim_{x\to+\infty}\varphi_{\scriptsize\mbox{K}}^{}(x) = 1.
\end{equation}
From Eq.~\eqref{eq:steqmo}, taking into account \eqref{eq:two}, one can easily obtain a first-order equation (also called the Bogomolny--Prasad--Sommerfield equation \cite{BPS1,BPS2})
\begin{equation}\label{eq:BPS}
    \frac{d\varphi}{dx} = \sqrt{2V(\varphi)}.
\end{equation}

Note that for a non-negative potential $V(\varphi)$ one can introduce a {\it superpotential} (sometimes called {\it prepotential}) --- a smooth function $W(\varphi)$ such that
\begin{equation}\label{eq:dwdfi}
    V(\varphi) = \frac{1}{2}\left(\frac{dW}{d\varphi}\right)^2.
\end{equation}
Then the energy \eqref{eq:stenergy} of a static kink (i.e., its {\it mass}) can be written as
\begin{equation}\label{eq:static_energy_bps}
   M_{\scriptsize\mbox{K}}^{} = \big|W(1)-W(-1)\big|.
\end{equation}
The mass of a part of the kink enclosed between the points $x_1^{}$ and $x_2^{}$ is obviously equal to $\big|W[\varphi(x_2^{})]-W[\varphi(x_1^{})]\big|$.

The kink excitation spectrum can be found as follows. A configuration $\varphi(x,t) = \varphi_{\scriptsize\mbox{K}}^{}(x) + \eta(x,t)$ with $|\eta| \ll |\varphi_{\scriptsize\mbox{K}}^{}|$ is substituted into the equation of motion \eqref{eq:eqmo}. Leaving only terms linear in $\eta(x,t)$, and searching for solution in the form $\eta(x,t) = \chi(x)\cos\:\omega t$, we get
\begin{equation}\label{eq:Shrodinger}
    \hat{H}\chi(x) = \omega^2\chi(x)
\end{equation}
with the Hamiltonian
\begin{equation}\label{eq:Schrod_Ham}
    \hat{H} = -\frac{d^2}{dx^2} + U(x),
\end{equation}
where
\begin{equation}\label{eq:Schrod_pot}
    U(x) = \left.\frac{d^2V}{d\varphi^2}\right|_{\varphi_{\rm K}^{}(x)}
\end{equation}
is called {\it stability potential} or {\it quantum-mechanical potential} of the kink. The function $\chi(x)$ is twice continuously differentiable and square-integrable on the entire real axis, $\lim\limits_{x\to\pm\infty}\chi(x)=0$, which corresponds to states of the discrete spectrum. Note that the same operator \eqref{eq:Schrod_Ham} can be obtained within the kink quantization in the second order term of the functional Taylor series for the energy \eqref{eq:stenergy} about $\varphi_{\scriptsize\mbox{K}}^{}(x)$, see, e.g., \cite[Ch.~5]{Rajaraman.book.1982}.

In the case of a kink defined implicitly as $x_{\scriptsize\mbox{K}}^{}(\varphi)$, the potential \eqref{eq:Schrod_pot} can only be obtained in the parametric form, i.e.
\begin{equation}\label{eq:stability_potential_parametric}
U(x): \quad
    \begin{cases}
        U(\varphi) = \displaystyle\frac{d^2V}{d\varphi^2},\\
        x(\varphi) = x_{\scriptsize\mbox{K}}^{}(\varphi).
    \end{cases}
\end{equation}
It can be easily shown that the discrete spectrum always has a zero level $\omega_0^{}=0$ (translational mode, see, e.g., \cite[Sec.~4]{Belendryasova.CNSNS.2019}), and the corresponding eigenfunction is $\chi_0^{}(x)=\displaystyle\frac{d\varphi_{\scriptsize\mbox{K}}^{}}{dx}$ or
\begin{equation}\label{eq:zero_mode_parametric}
\chi_0^{}(x): \quad
    \begin{cases}
        \chi_0^{}(\varphi) = \displaystyle\frac{1}{x_{\scriptsize\mbox{K}}^\prime(\varphi)},\\
        x(\varphi) = x_{\scriptsize\mbox{K}}^{}(\varphi).
    \end{cases}
\end{equation}
For a kink with (at least one) power-law asymptotics, the zero level is located at the boundary of the continuous spectrum, the wave function of the zero mode $\chi_0^{}(x)$ has no nodes, and in the case of a power-law kink asymptotics at $x\to+\infty$ (at $x\to-\infty$) decreases power-law as $x\to+\infty$ (as $x\to-\infty$). There are clearly no other levels in the discrete part of the kink's excitation spectrum, except for $\omega_0^{}=0$.

\section{Kinks in a class of polynomial models}
\label{sec:Solutions}

Consider a class of models with potentials
\begin{equation}\label{eq:potential_general}
    V(\varphi) = \frac{1}{2}\left(1+\varphi\right)^{2m}\left(1-\varphi\right)^{2n},
\end{equation}
where $m$ and $n$ are positive integers. If $m=1$ ($n=1$) then the kink asymptotics at $\varphi\to -1$ ($\varphi\to 1$) is exponential. If $m\ge 2$ ($n\ge 2$) then the kink asymptotics at $\varphi\to -1$ ($\varphi\to 1$) is power-law. At $m=n=1$, Eq.~\eqref{eq:potential_general} gives the potential of the well-known $\varphi^4$ model \cite{Kevrekidis.book.2019,Belova.UFN.1997} with kink $\varphi_{\scriptsize\mbox{K}}^{}(x)=\tanh x$. We will not consider this case.

Kinks with power-law asymptotics for potentials \eqref{eq:potential_general} can only be obtained in an implicit form as $x=x_{\scriptsize\mbox{K}}^{}(\varphi)$. Accordingly, both the stability potential and the zero mode eigenfunction can only be obtained in a parametric form. 

Let's consider kinks corresponding to \eqref{eq:potential_general} potentials, as well as their properties, in more detail. We start with two special cases, and then move on to the most general case.

\subsection{Special case $m=1$ and $n\ge 2$}
\label{sec:Case_A}

\begin{figure}[t!]
    \centering
    \includegraphics[width=0.6\textwidth]{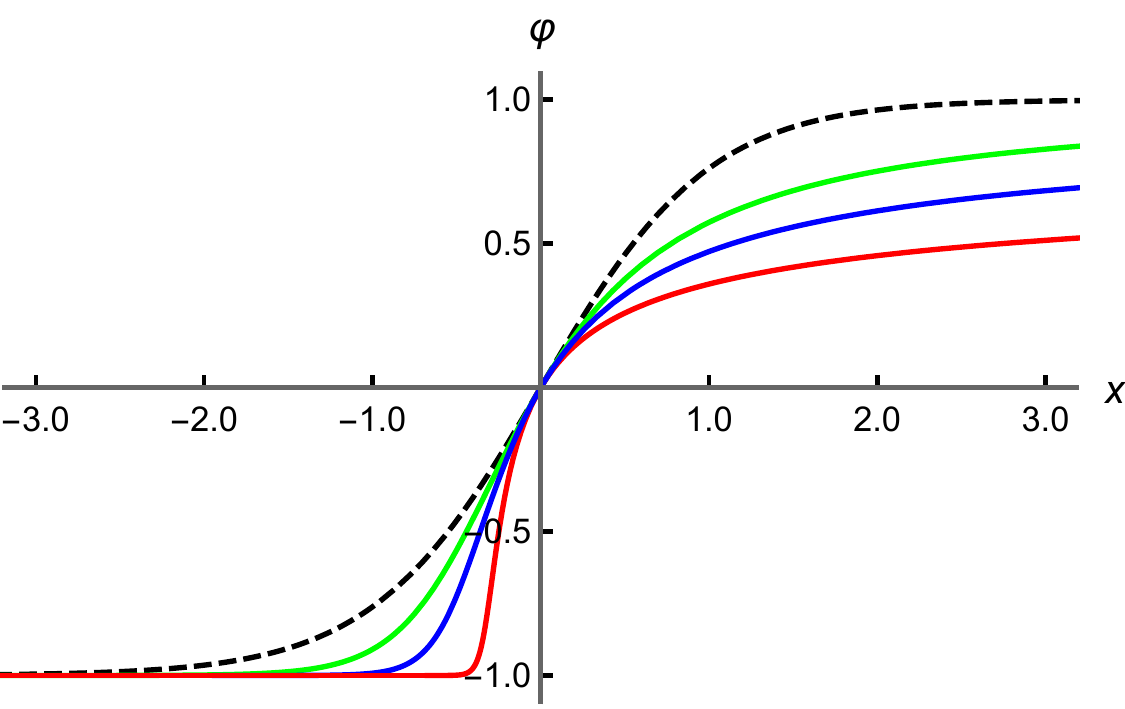}
    \caption{Kinks \eqref{eq:kink_for_m_equals_1} for $n=2$ (green curve), $n=3$ (blue curve), $n=5$ (red curve). For comparison, the $\varphi^4$ kink is shown (black dashed curve), which corresponds to $n=1$ in the potential \eqref{eq:potential_for_m_equals_1}.}
    \label{fig:kink_for_m_equals_1}
\end{figure}

For $m=1$ and $n\ge 2$ the potential \eqref{eq:potential_general} looks like
\begin{equation}\label{eq:potential_for_m_equals_1}
    V(\varphi) = \frac{1}{2}\left(1+\varphi\right)^2\left(1-\varphi\right)^{2n}.
\end{equation}
Integrating the ordinary differential equation \eqref{eq:BPS} with the potential \eqref{eq:potential_for_m_equals_1}, we obtain the kink solution
\begin{equation}\label{eq:kink_for_m_equals_1}
    x - x_0^{}= \frac{1}{2^n}\ln\frac{1+\varphi}{1-\varphi} + \frac{1}{2^n}\sum_{j=1}^{n-1}\frac{2^j}{j}\frac{1}{\left(1-\varphi\right)^j},
\end{equation}
see Fig.~\ref{fig:kink_for_m_equals_1}.
It is convenient to choose the constant $x_0^{}$ in Eq.~\eqref{eq:kink_for_m_equals_1} so that $\varphi=0$ for $x=0$: $x_0^{}=-\displaystyle\frac{1}{2^n}\sum\limits_{j=1}^{n-1}\frac{2^j}{j}$.

The kink \eqref{eq:kink_for_m_equals_1} has left exponential asymptotics and right power-law asymptotics:
\begin{equation}\label{eq:kink_for_m_equals_1_asymptotics}
    \varphi_{\scriptsize\mbox{K}}^{}(x) \approx
    \begin{cases}
    -1 + 2\exp\left[2^n (x-x_0^{})\right] \quad \mbox{at} \quad x\to-\infty,\vspace{1mm}\\
    \thinspace\thinspace\thinspace\thinspace 1 - \displaystyle\frac{\left[2(n-1)\right]^{1/(1-n)}}{(x-x_0^{})^{1/(n-1)}} \quad\thinspace\thinspace\thinspace\thinspace \mbox{at} \quad x\to+\infty,
    \end{cases}
\end{equation}
for more details see, e.g., \cite[Sec.~4]{Blinov.AoP.2022}, \cite[Sec.~3]{Blinov.JPCS.2020.deform}. Note that the pre-exponential factor in \eqref{eq:kink_for_m_equals_1_asymptotics} does not carry any deep physical meaning, it depends on which point is considered to be the location of the kink.

Analyzing Eq.~\eqref{eq:kink_for_m_equals_1_asymptotics}, one can also notice that as $n$ increases, the rate of exponential asymptotics at $x\to-\infty$ (the coefficient in front of $x$ in the exponent) increases, while the rate of power-law asymptotics at $x \to+\infty$ decreases.

The superpotential (up to an additive constant) corresponding to the potential \eqref{eq:potential_for_m_equals_1} is
\begin{equation}\label{eq:superpotential_for_m_equals_1}
    W(\varphi) = - \frac{\left(1-\varphi\right)^{n+1}\left[\left(n+1\right)\varphi+n+3\right]}{(n+1)(n+2)}.
\end{equation}
Then the mass of the kink \eqref{eq:kink_for_m_equals_1} can be found from Eq.~\eqref{eq:static_energy_bps}:
\begin{equation}\label{eq:mass_for_m_equals_1}
    M_{\scriptsize\mbox{K}}^{} = \frac{2^{n+2}}{(n+1)(n+2)}.
\end{equation}
It is seen that $M_{\scriptsize\mbox{K}}^{}\to +\infty$ as $n\to\infty$.

\begin{figure}[t!]
    \centering
    \includegraphics[width=0.6\textwidth]{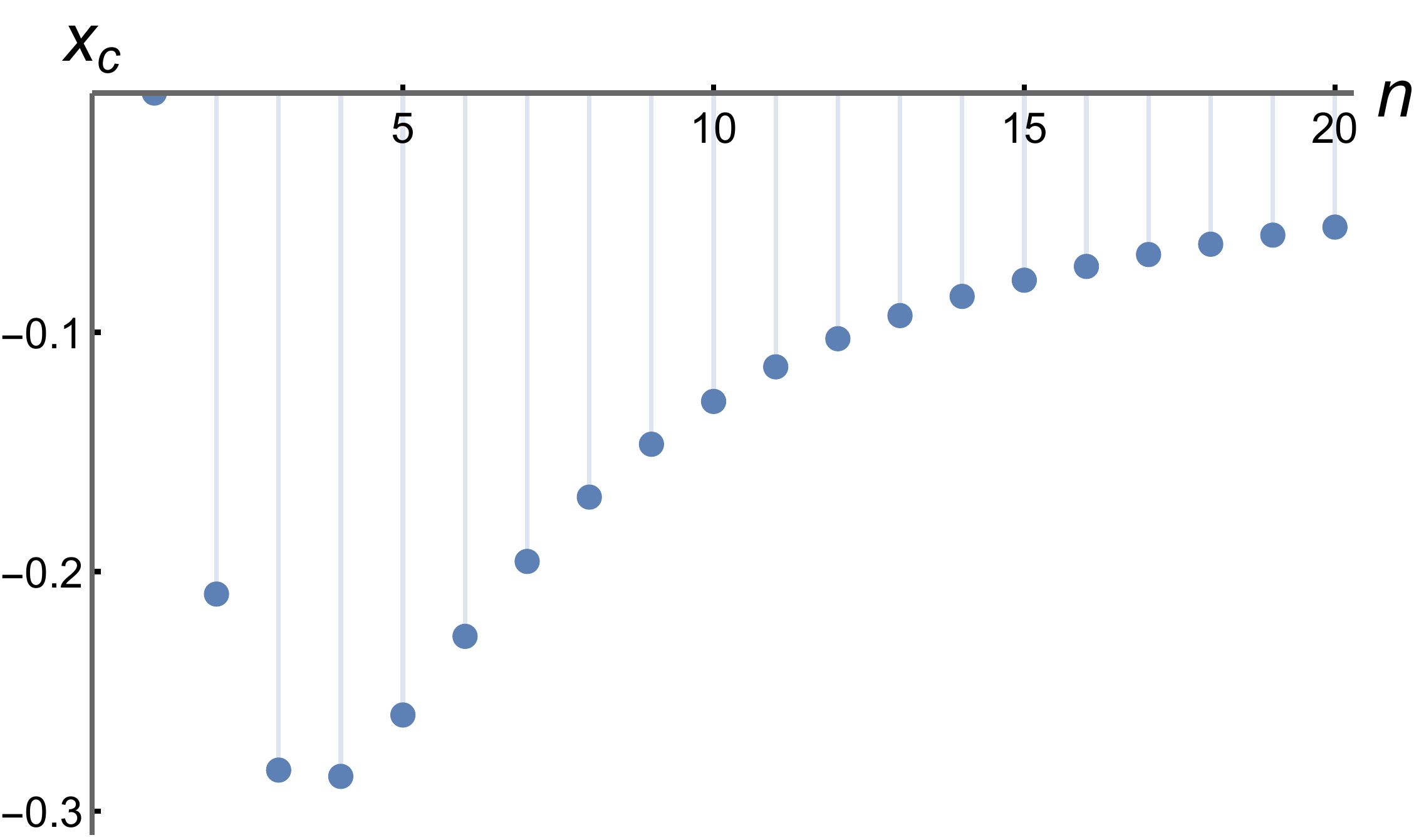}
    \caption{Dependence of the coordinate $x_{\rm c}^{}$ of the center of mass of the kink \eqref{eq:kink_for_m_equals_1} on the parameter $n$ of the potential \eqref{eq:potential_for_m_equals_1}.}
    \label{fig:com_of_n_m_eq_1}
\end{figure}

The position of the center of mass of the kink \eqref{eq:kink_for_m_equals_1} can be obtained, for example, from the condition that equal parts of the kink mass are located to the left and to the right of the center of mass. Taking into account Eqs.~\eqref{eq:superpotential_for_m_equals_1} and \eqref{eq:mass_for_m_equals_1}, for $\varphi_{\rm c}^{}$ corresponding to the center of mass we obtain the algebraic equation
\begin{equation}
    \left(1-\varphi_{\rm c}^{}\right)^{n+1}\left[(n+1)\varphi_{\rm c}^{}+n+3\right] = 2^{n+1}.
\end{equation}
To find the coordinate $x_{\rm c}^{}$ of the center of mass, we need to substitute $\varphi_{\rm c}^{}$ into the expression for the kink \eqref{eq:kink_for_m_equals_1}.
The dependence $x_{\rm c}^{}(n)$ is shown in Fig.~\ref{fig:com_of_n_m_eq_1}.

\begin{figure}[t!]
    \centering
    \includegraphics[width=0.8\textwidth]{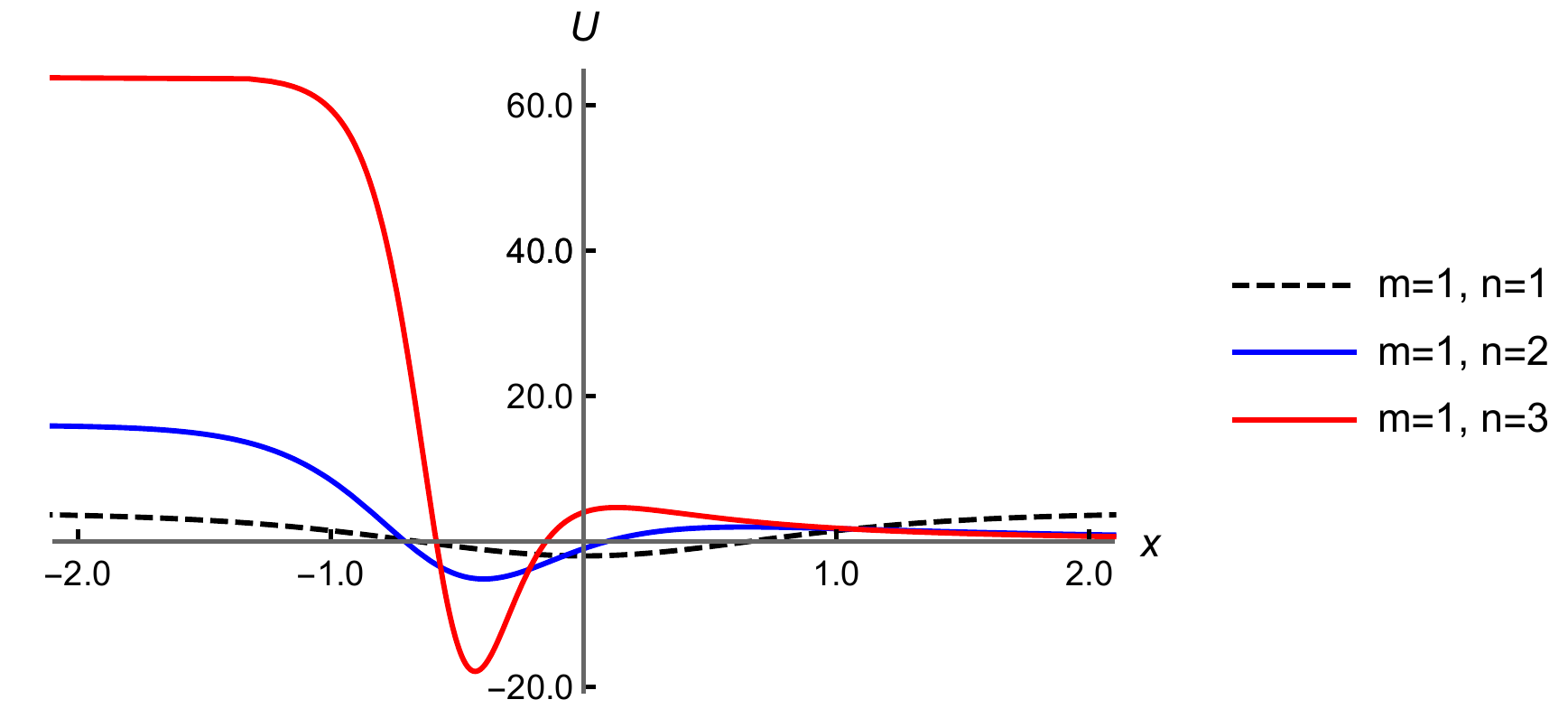}
    \caption{Stability potential \eqref{eq:stability_potential_parametric} for the kink \eqref{eq:kink_for_m_equals_1}; for comparison, the stability potential of the $\varphi^4$ kink is shown, which corresponds to $m=1$, $n=1$ in the potential \eqref{eq:potential_general}.}
    \label{fig:quantum_mechanical_potential_m_1}
\end{figure}

The stability potential for the kink \eqref{eq:kink_for_m_equals_1} is found numerically form Eq.~\eqref{eq:stability_potential_parametric}. In Fig.~\ref{fig:quantum_mechanical_potential_m_1},
several cases are presented for different values of $n$. The zero mode eigenfunction, which is the derivative of the kink \eqref{eq:kink_for_m_equals_1} up to a normalization factor, can be easily found from Eqs.~\eqref{eq:zero_mode_parametric} and \eqref{eq:kink_for_m_equals_1}.

\subsection{Special case $m=n\ge 2$}
\label{sec:Case_B}

\begin{figure}[t!]
    \centering
    \includegraphics[width=0.6\textwidth]{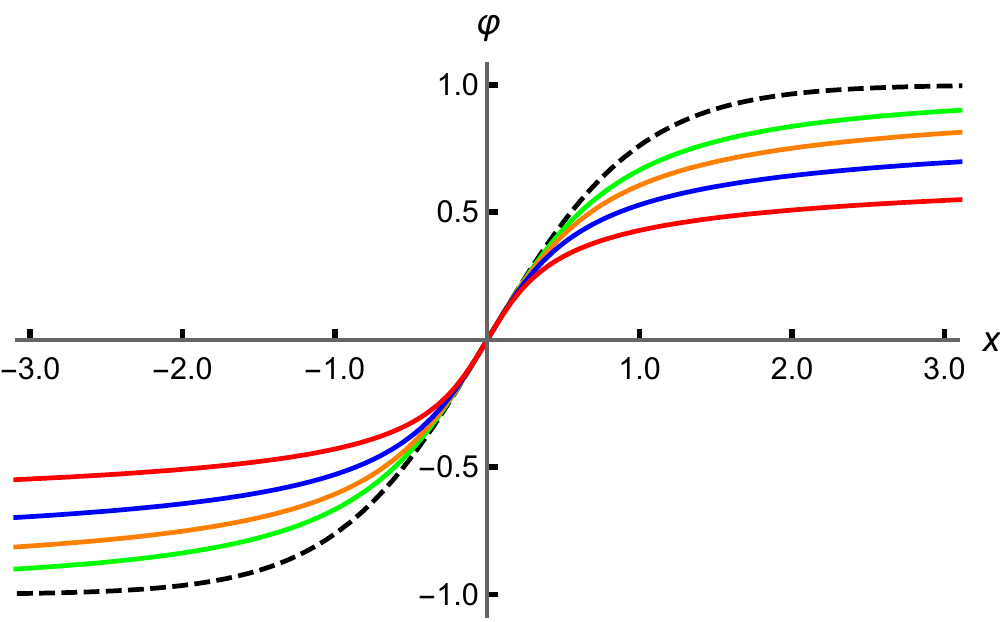}
    \caption{Kinks \eqref{eq:kink_for_m_equals_n} for $n=2$ (green curve), $n=3$ (orange curve), $n=5$ (blue curve), $n=10$ (red curve). For comparison, the $\varphi^4$ kink is shown (black dashed curve), which corresponds to $n=1$ in the potential \eqref{eq:potential_for_m_equals_n}.}
    \label{fig:kink_for_m_equals_n}
\end{figure}

For $m=n\ge 2$ the potential \eqref{eq:potential_general} is
\begin{equation}\label{eq:potential_for_m_equals_n}
    V(\varphi) = \frac{1}{2}\left(1-\varphi^2\right)^{2n},
\end{equation}
and the corresponding kinks look like
\begin{equation}\nonumber
    x = \frac{C_{2n-3}^{n-1}}{8} \cdot \varphi \cdot \sum_{j=0}^{n-3}\frac{1}{4^j\left(n-j-2\right)C_{2n-2j-5}^{n-j-2}} \cdot \frac{1}{\left(1-\varphi^2\right)^{n-j-2}} +
\end{equation}
\begin{equation}\label{eq:kink_for_m_equals_n}
    + \frac{1}{2\left(n-1\right)}\cdot\frac{\varphi}{\left(1-\varphi^2\right)^{n-1}} - \frac{C_{2n-3}^{n-1}}{4^{n-1}}\ln\frac{1-\varphi}{1+\varphi},
\end{equation}
see Fig.~\ref{fig:kink_for_m_equals_n}.
The kinks \eqref{eq:kink_for_m_equals_n} are centered so that $\varphi=0$ at $x=0$, they are symmetric and have power-law asymptotics, that can be obtained either by analyzing the potential \eqref{eq:potential_for_m_equals_n} or from Eq.~\eqref{eq:kink_for_m_equals_n}:
\begin{equation}\label{eq:asymptotics_for_m_equals_n}
    \varphi_{\scriptsize\mbox{K}}^{}(x) \approx \pm 1 \mp \frac{\left[2^n(n-1)\right]^{1/(1-n)}}{|x|^{1/(n-1)}} \quad \mbox{at} \quad x\to\pm\infty.
\end{equation}

Superpotential corresponding to the potential \eqref{eq:potential_for_m_equals_n} has the form
\begin{equation}\label{eq:superpotential_for_m_equals_n}
    W(\varphi) = \sum\limits_{j=0}^n \frac{(-1)^j C_n^j}{2j+1} \cdot \varphi^{2j+1}.
\end{equation}
Using it, we get the mass of the kink \eqref{eq:kink_for_m_equals_n}:
\begin{equation}\label{eq:mass_for_m_equals_n}
    M_{\scriptsize\mbox{K}}^{} = 2\sum\limits_{j=0}^n \frac{(-1)^j C_n^j}{2j+1} = \frac{2\cdot(2n)!!}{(2n+1)!!}.
\end{equation}
From Eq.~\eqref{eq:mass_for_m_equals_n} it follows that $M_{\scriptsize\mbox{K}}^{}\to 0$ as $n\to\infty$. The calculation of the sum is described in the Appendix \ref{sec:Appendix_A}. Note that the mass of the kink can also be found from Eq.~\eqref{eq:stenergy} using Eq.~\eqref{eq:BPS}, which leads to an integral that reduces to the Euler beta function, see Appendix \ref{sec:Appendix_D}.

Note that our expression for the kink \eqref{eq:kink_for_m_equals_n} coincides (taking into account the difference in the choice of model parameters) with the one obtained in Ref.~\cite[Sec.~3]{Khare.JPA.2019}. On the other hand, our formula \eqref{eq:mass_for_m_equals_n} for the mass differs from that obtained in Ref.~\cite[Sec.~3]{Khare.JPA.2019}. Interestingly, the two masses coincide at $n=2$, and differ for $n\ge 3$; in the limit $n\to\infty$, the ratio of the two masses (our to their) is $\sqrt{2}$.

\begin{figure}[t!]
    \centering
    \includegraphics[width=0.8\textwidth]{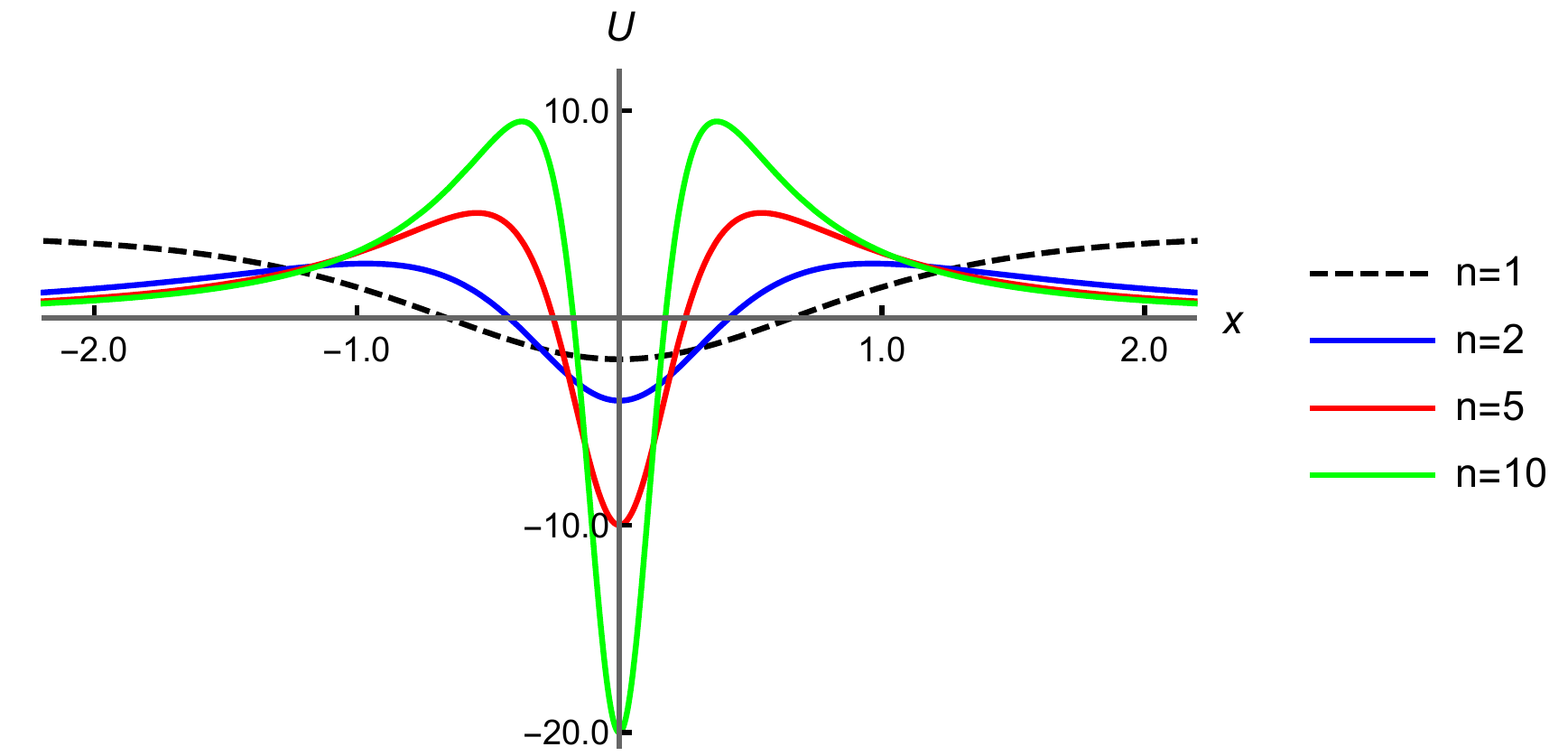}
    \caption{Stability potential \eqref{eq:stability_potential_parametric} for the kink \eqref{eq:kink_for_m_equals_n}; for comparison, the stability potential of the $\varphi^4$ kink is shown, which corresponds to $n=1$ in the potential \eqref{eq:potential_for_m_equals_n}.}
    \label{fig:quantum_mechanical_potential_m_eq_n}
\end{figure}

Since the kinks \eqref{eq:kink_for_m_equals_n} are symmetric, their centers of mass are located at the origin. In Fig.~\ref{fig:quantum_mechanical_potential_m_eq_n}
the stability potential is shown for several $n$. In all cases, the potential is symmetric and volcano-like. The discrete spectrum contains only zero mode, the corresponding eigenfunction is symmetric and can be easily found from Eqs.~\eqref{eq:zero_mode_parametric} and \eqref{eq:kink_for_m_equals_n}.

\subsection{General case of arbitrary $m\ge 2$ and $n\ge 2$}
\label{sec:Case_C}

\begin{figure}[t!]
    \centering
    \includegraphics[width=0.6\textwidth]{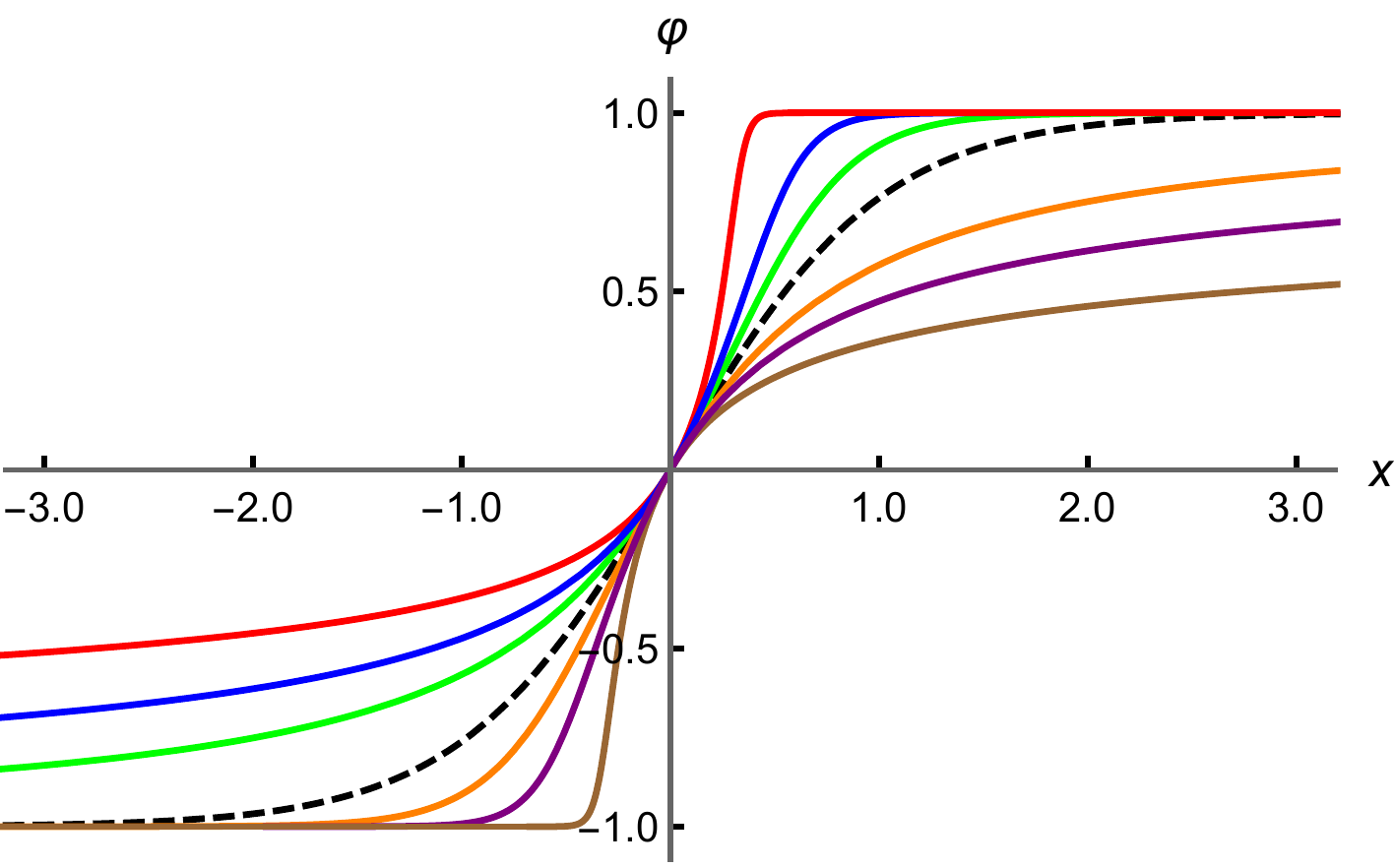}
    \caption{Kinks \eqref{eq:kink_general} for $m=5$, $n=1$ (red curve), $m=3$, $n=1$ (blue curve), $m=2$, $n=1$ (green curve), $m=1$, $n=2$ (orange curve), $m=1$, $n=3$ (purple curve), $m=1$, $n=5$ (brown curve). For comparison, the $\varphi^4$ kink is shown (black dashed curve), which corresponds to $m=1$, $n=1$ in the potential \eqref{eq:potential_general}.}
    \label{fig:kink_for_general_potential}
\end{figure}

For the potential \eqref{eq:potential_general} with arbitrary $m\ge 2$ and $n\ge 2$ we obtain the kink
\begin{equation}\nonumber
    x - x_0^{} = -\frac{1}{(1-\varphi)^{n-1}}\sum\limits_{j=1}^{m-1} \frac{C_{m+n-2}^{j-1}}{2^j\left(m-j\right)C_{m-1}^{j-1}} \cdot \frac{1}{(1+\varphi)^{m-j}} +
\end{equation}
\begin{equation}\label{eq:kink_general}
    + \frac{C_{m+n-2}^{m-1}}{2^{m-1}} \sum\limits_{l=1}^{n-1}  \frac{1}{2^l\left(n-l\right)} \cdot\frac{1}{(1-\varphi)^{n-l}} + \frac{C_{m+n-2}^{m-1}}{2^{m+n-1}}\ln\frac{1+\varphi}{1-\varphi},
\end{equation}
see Fig.~\ref{fig:kink_for_general_potential}.
It is convenient to choose the constant $x_0^{}$ in Eq.~\eqref{eq:kink_general} so that $\varphi=0$ for $x=0$: $x_0^{}=\displaystyle\sum\limits_{j=1}^{m-1}\frac{C_{m+n-2}^{j-1}}{2^j\left(m-j\right)C_{m-1}^{j-1}}-\frac{C_{m+n-2}^{m-1}}{2^{m-1}}\sum\limits_{l=1}^{n-1}\frac{1}{2^l(n-l)}$. The asymptotics of the solution \eqref{eq:kink_general} can be obtained by analyzing the potential \eqref{eq:potential_general}:
\begin{equation}\label{eq:asymptotics_general}
    \varphi_{\scriptsize\mbox{K}}^{}(x) \approx \begin{cases}
    -1 + \displaystyle\frac{\left[2^n(m-1)\right]^{1/(1-m)}}{|x-x_0^{}|^{1/(m-1)}} \quad \mbox{at} \quad x\to-\infty,\\
    \thinspace \thinspace \thinspace \thinspace \thinspace 1 - \displaystyle\frac{\left[2^m(n-1)\right]^{1/(1-n)}}{(x-x_0^{})^{1/(n-1)}} \quad \thinspace \mbox{at} \quad x\to+\infty.
    \end{cases}
\end{equation}
On the other hand, the same asymptotics can be extracted directly from Eq.~\eqref{eq:kink_general}:
\begin{equation}
    \varphi_{\scriptsize\mbox{K}}^{}(x) \approx \begin{cases}
    -1 + \displaystyle\frac{\left[2^n(m-1)\right]^{1/(1-m)}}{|x-x_0^{}|^{1/(m-1)}} \quad \mbox{at} \quad x\to-\infty,\\
    \thinspace \thinspace \thinspace \thinspace \thinspace 1 - \displaystyle\frac{A^{1/(n-1)}}{(x-x_0^{})^{1/(n-1)}} \quad \quad \thinspace \thinspace \thinspace \mbox{at} \quad x\to+\infty,
    \end{cases}
\end{equation}
where
\begin{equation}\label{eq:A}
    A = \frac{1}{2^m} \left[ \frac{C_{m+n-2}^{m-1}}{n-1} - \displaystyle\sum\limits_{j=1}^{m-1}\frac{C_{m+n-2}^{j-1}}{\left(m-j\right)C_{m-1}^{j-1}} \right].
\end{equation}
The sum in this formula can be summed, resulting in Eq.~\eqref{eq:asymptotics_general}, see Appendix \ref{sec:Appendix_B}.

Superpotential corresponding to the potential \eqref{eq:potential_general} reads
\begin{equation}\label{eq:superpotential_general}
    W(\varphi) = \sum\limits_{j=0}^m\sum\limits_{l=0}^n\frac{(-1)^l C_m^j C_n^l}{j+l+1} \cdot \varphi^{j+l+1},
\end{equation}
and mass of the kink \eqref{eq:kink_general} is
\begin{equation}
    M_{\scriptsize\mbox{K}}^{} = \sum\limits_{j=0}^m\sum\limits_{l=0}^n\frac{\left[(-1)^j+(-1)^l\right]C_m^j C_n^l}{j+l+1}.
\end{equation}
This sum can be calculated as shown in Appendix \ref{sec:Appendix_C}. On the other hand, the mass of the kink can be calculated from Eq.~\eqref{eq:stenergy} using Eq.~\eqref{eq:BPS}, see Appendix \ref{sec:Appendix_D} for details. As a result, we get
\begin{equation}\label{eq:mass_general}
    M_{\scriptsize\mbox{K}}^{} = \frac{2^{m+n+1}}{m+n+1} \cdot \frac{1}{C_{m+n}^{n}}.
\end{equation}

\begin{figure}[t!]
\begin{minipage}[h]{0.49\linewidth}
\center{\includegraphics[width=0.9\textwidth]{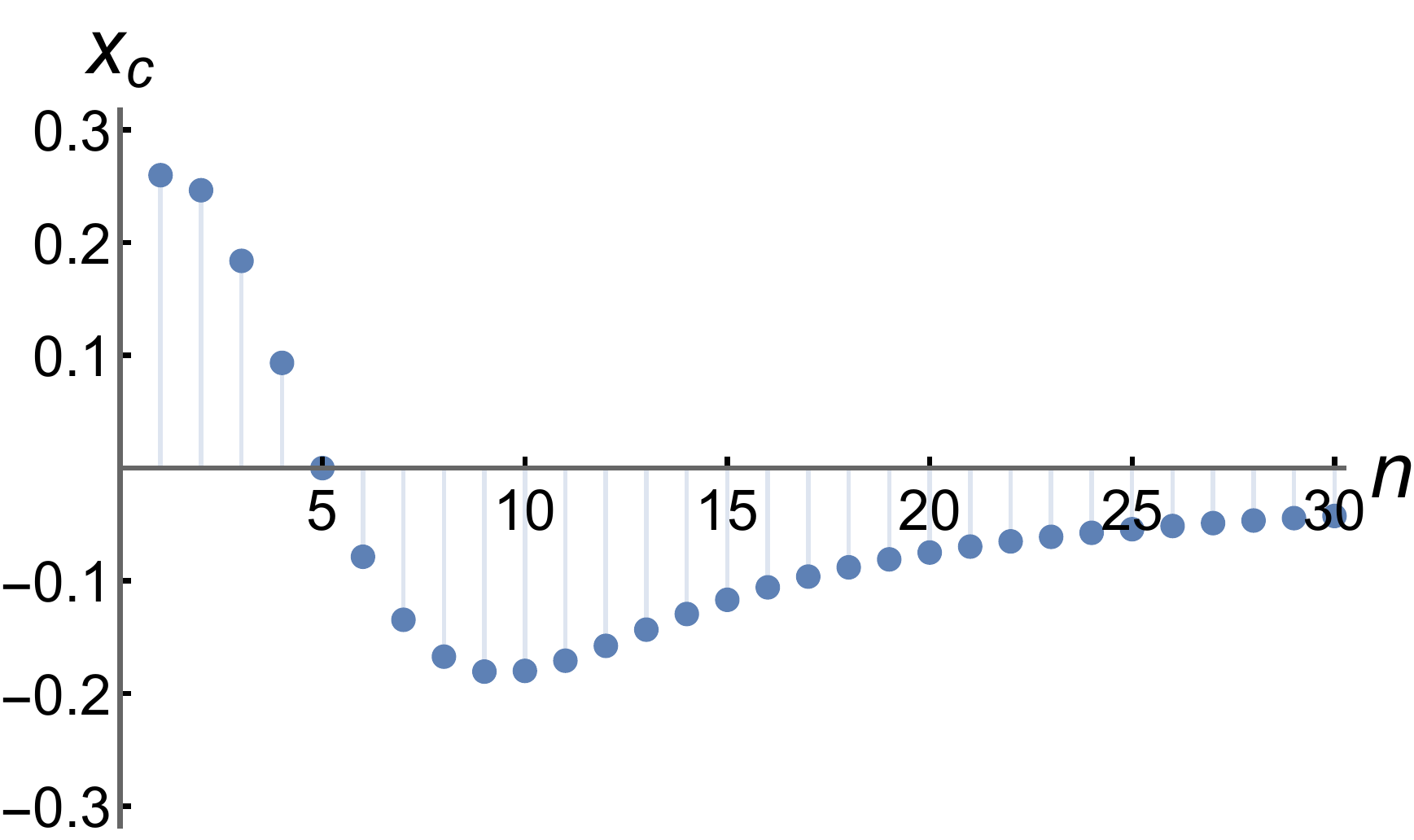} \\ (a) $m = 5$}
\end{minipage}
\hfill
\begin{minipage}[h]{0.49\linewidth}
\center{\includegraphics[width=0.9\textwidth]{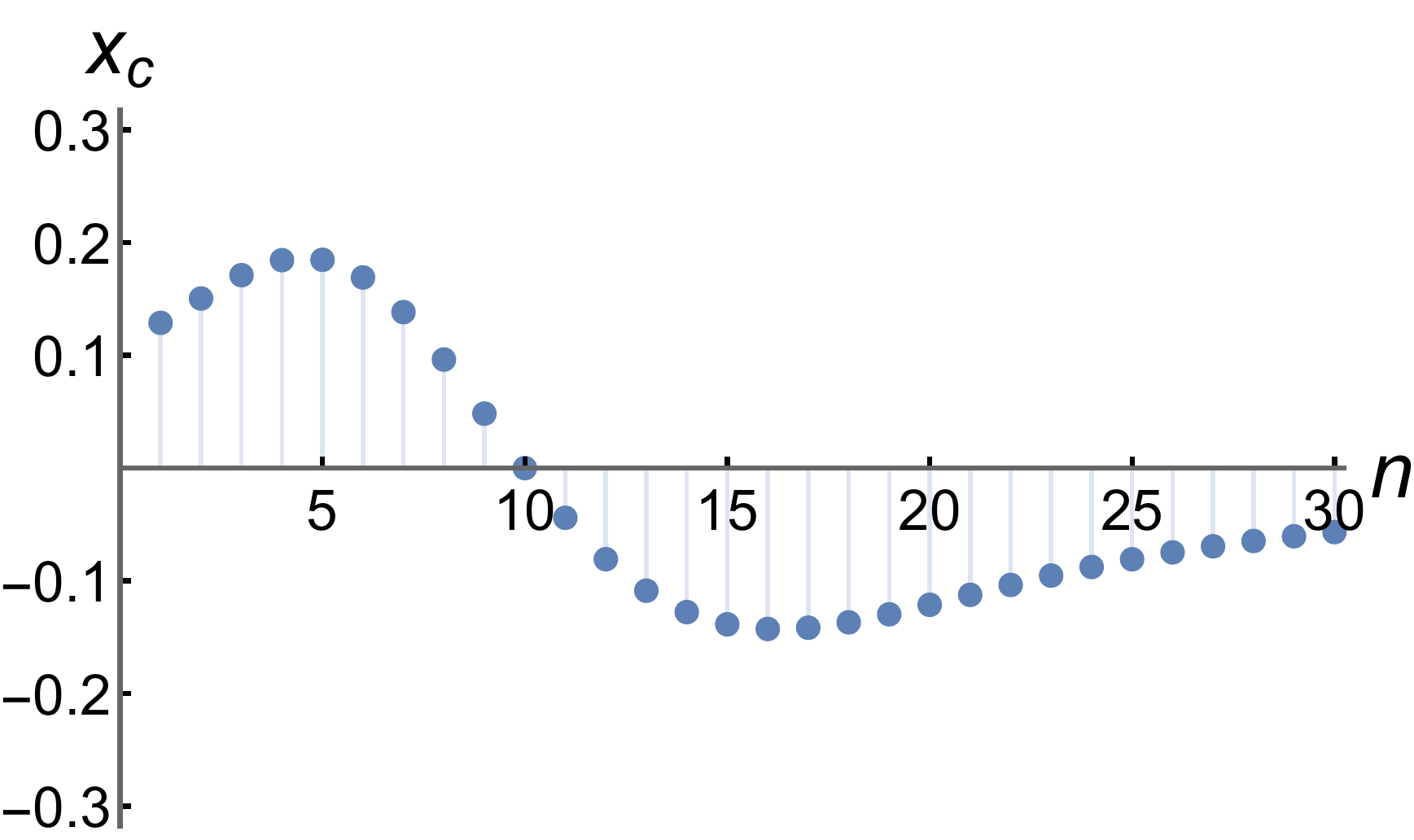} \\ (b) $m = 10$}
\end{minipage}
\caption{Position of the center of mass of the kink \eqref{eq:kink_general} as a function of $n$ for two different $m$.}
\label{fig:com_of_n_for_different_m}
\end{figure}

The position of the center of mass of the kink \eqref{eq:kink_general} can be found in the same way as it was done in Section \ref{sec:Case_A}. Taking into account Eq.~\eqref{eq:superpotential_general} and Eq.~\eqref{eq:mass_general}, for $\varphi_{\rm c}^{}$ we obtain the algebraic equation
\begin{equation}
  \sum\limits_{j=0}^m\sum\limits_{l=0}^n\frac{(-1)^l C_m^j C_n^l}{j+l+1} \cdot \varphi_{\rm c}^{j+l+1} = \frac{1}{2}\left( \sum\limits_{j=0}^m \frac{m!\:n!}{(m-j)!\:(j+n+1)!}-\sum\limits_{l=0}^n \frac{m!\:n!}{(n-l)!\:(l+m+1)!}\right).
\end{equation}
Substituting the value of $\varphi_{\rm c}^{}$ found (numerically) from here into Eq.~\eqref{eq:kink_general}, we get $x_{\rm c}^{}$. The obtained dependencies $x_{\rm c}^{}(n)$ are shown in Fig.~\ref{fig:com_of_n_for_different_m} for $m=5$ and $m=10$. As $m$ increases, the local maximum and local minimum shift to the right.

\begin{figure}[t!]
    \centering
    \includegraphics[width=0.8\textwidth]{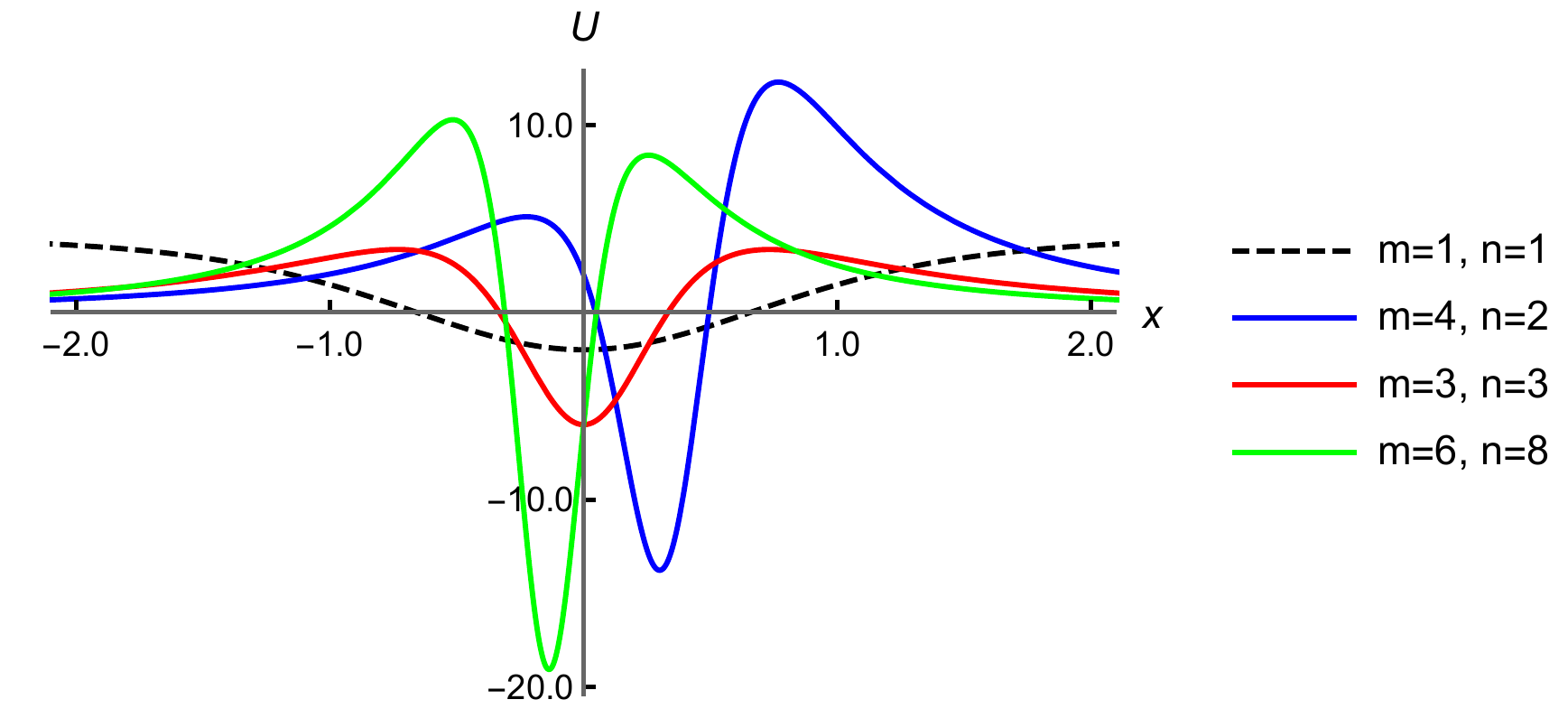}
    \caption{Stability potential \eqref{eq:stability_potential_parametric} for the kink \eqref{eq:kink_general}, for comparison, the stability potential of the $\varphi^4$ kink is shown, which corresponds to $m=1$, $n=1$ in the potential \eqref{eq:potential_general}.}
    \label{fig:quantum_mechanical_potential_general}
\end{figure}

The stability potential of the kink \eqref{eq:kink_general} is shown in Fig.~\ref{fig:quantum_mechanical_potential_general}.
For all $m\ge 2$, $n\ge 2$ the potential $U(x)$ is volcano-like. Besides that, as was shown in Ref.~\cite{Blinov.AoP.2022}, in the case of power-law asymptotics of the kink we always have $U(x)\to 0+0$, $U(x)\sim 1/x^2$ as $x\to\pm\infty$. The discrete spectrum in the potential well $U(x)$ contains only zero mode, the corresponding eigenfunction is asymmetric and can be found from Eqs.~\eqref{eq:zero_mode_parametric} and \eqref{eq:kink_general}.

\section{Conclusion}
\label{sec:Conclusion}

We have considered a family of field-theoretic models with polynomial potentials $V(\varphi) = \displaystyle\frac{1}{2}\left(1+\varphi\right)^{2m}\left(1-\varphi\right)^{2n}$, $m\ge 1$, $n\ge 2$. We obtained solutions of the type of topological solitons (kink solutions), as well as studied their asymptotic behavior, which is important for various physical applications.

For all kinks, we obtained the stability potentials, which determine the excitation spectra, as well as interactions of kinks with scalar radiation --- small-amplitude waves with frequencies lying in the region $\omega^2>0$ (for the discussed family of models). We also found the zero mode eigenfunctions in all cases.

For $m=n$ the solutions found are (as expected) symmetric, while for $m\neq n$ the kinks are asymmetric. In the asymmetric case, the question of the mass distribution along the soliton is nontrivial. For all kinks, we have found the position of the center of mass. Interestingly, at $m=1$, as $n$ increases, the center of mass first shifts to the left, the shift reaches its maximum at $n=4$. As $n$ increases further, the center of mass gradually approaches $x=0$ (recall that in all cases we centered the kinks so that $\varphi=0$ at $x=0$). Analyzing the case of arbitrary $m$ and $n$, we see that the center of mass is shifted to the left for $m<n$ and to the right for $m>n$.

It is noteworthy that, up to a factor, the considered potentials coincide with the squares of the primary Bernstein polynomials on the segment $[-1,1]$, see \cite[Eq.~(14)]{Petrosova.VMJ.2019}. This, in turn, immediately leads to the fact that when calculating the kink mass, we obtain combinatorial structures that form the so-called {\it Pascal trapezoids}, recently introduced in Ref.~\cite{Petrosova.VMJ.2019}. Related objects are actively used in probability theory and cryptography as the values of special {\it Kravchuk polynomials} (also known as {\it Krawtchouk polynomials}) \cite{Feinsilver.book.2005}.

The obtained general formulas for topological solitons can be further used to solve many problems. In particular:
\begin{itemize}
    \item Interaction forces between two or more solitons can be found.
    (In the case of large distances, one should use the asymptotic expressions also obtained in our paper.) For such calculations, the methods proposed in \cite{Christov.PRL.2019,Manton.JPA.2019} can be used. We hope to publish our results of such estimations soon.
    \item As already mentioned in Section \ref{sec:Introduction}, resonance phenomena were previously observed in the collision of solitons with power-law asymptotics \cite{Belendryasova.CNSNS.2019,Christov.CNSNS.2021}. Verification of the mechanism proposed in Ref.~\cite{Belendryasova.CNSNS.2019}, or the construction of a more accurate quantitative theory describing the resonant exchange between kinetic energy and some energy accumulator, is an important problem awaiting its solution. In this context, knowledge of the stability potential that determines the kink's excitation spectrum is necessary, at least in the implicit form.
\end{itemize}

\section*{Acknowledgments}

This work was supported by MEPhI within the Program ``Priority-2030'' under the contract No.~075-15-2021-1305.

\appendix

\section{}
\label{sec:Appendix_A}

Let $n\in\mathbb{N}\cup\{0\}$. We will demonstrate that
\begin{equation}\nonumber
    \sum\limits_{j=0}^n \frac{(-1)^j C_n^j}{2j+1} = \frac{(2n)!!}{(2n+1)!!} = \frac{1}{2n+1}\frac{2^{2n}}{C_{2n}^n}.
\end{equation}
Consider the polynomial
\begin{equation}\nonumber
    S_n^{}(x) = \sum\limits_{j=0}^n \frac{(-1)^j C_n^j}{2j+1}\: x^{2j+1}.
\end{equation}
For this polynomial we have $S_n^{}(0)=0$ and
\begin{equation}\nonumber
    S_n^\prime(x) = \sum\limits_{j=0}^n (-1)^j C_n^j x^{2j} = \left(1-x^2\right)^n.
\end{equation}
Hence, $S_n^{}(x)=\displaystyle\int_0^x \left(1-t^2\right)^ndt$, and the sum we are looking for is equal to $S_n^{}(1)$. By integrating by parts, one can obtain the recurrence formula
\begin{equation}\nonumber
    S_n^{}(1) = \frac{2n}{2n+1} S_{n-1}^{}(1), \quad \mbox{where} \quad S_0^{}(1) = 1.
\end{equation}
Applying successively obtained relation, we find that
\begin{equation}\nonumber
    S_n^{}(1) = \frac{2n}{2n+1} \cdot S_{n-1}^{}(1) = \frac{2n}{2n+1} \cdot \frac{2n-2}{2n-1} \cdot S_{n-2}^{}(1) = ... = \frac{(2n)!!}{(2n+1)!!}.
\end{equation}

\section{}
\label{sec:Appendix_B}

Let us show how the sum $\displaystyle\sum\limits_{j=1}^{m-1}\frac{C_{m+n-2}^{j-1}}{(m-j)C_{m-1}^{j-1}}$ in Eq.~\eqref{eq:A} can be calculated. First do the following transformations:
\begin{equation}\nonumber
    \frac{C_{m+n-2}^{j-1}}{(m-j)\:C_{m-1}^{j-1}} = \frac{(m+n-2)!\:\cancel{(j-1)!}\:\cancelto{(m-j-1)!}{(m-j)!}}{\cancel{(j-1)!}\:\cancel{(m-j)}\:(m+n-j-1)!\:(m-1)!} =
\end{equation}
\begin{equation}\label{eq:B1}
    = \underbrace{\frac{(m+n-2)!}{(m-1)!\:(n-1)!}}_{C_{m+n-2}^{m-1}} \cdot \underbrace{\frac{n!\:(m-j-1)!}{(m+n-j-1)!}}_{1/C_{m+n-j-1}^n} \cdot\: \frac{1}{n} = \frac{C_{m+n-2}^{m-1}}{n} \cdot \frac{1}{C_{m+n-j-1}^n}.
\end{equation}
Then the original sum can be reduced to the form
\begin{equation}\label{eq:B2}
    \frac{C_{m+n-2}^{m-1}}{n} \sum\limits_{j=1}^{m-1}\frac{1}{C_{m+n-j-1}^n} = \frac{C_{m+n-2}^{m-1}}{n} \sum\limits_{j=1}^k\frac{1}{C_{p-j}^{p-k}} = \frac{C_{m+n-2}^{m-1}}{n} \cdot \sigma_p^{}(k),
\end{equation}
where the indices are replaced in accordance with $m-1=k$ and $n+k=p$, and the sum is denoted by $\sigma_p^{}(k)$. Now we calculate this sum:
\begin{equation}\nonumber
    \sigma_p^{}(k+1) = \sum\limits_{j=1}^{k+1}\frac{1}{C_{p-j}^{p-k-1}} = \sum\limits_{j=1}^k\frac{(p-j)-(p-k-1)}{k+1-j}\cdot\frac{1}{C_{p-j}^{p-k-1}} + 1 =
\end{equation}
\begin{equation}\nonumber
    = \sum\limits_{j=1}^k\underbrace{\frac{p-j}{k+1-j}\cdot\frac{1}{C_{p-j}^{p-k-1}}}_{1/C_{p-j-1}^{p-k-1}} - \sum\limits_{j=1}^k\underbrace{\frac{p-k-1}{k+1-j}\cdot\frac{1}{C_{p-j}^{p-k-1}}}_{\frac{p-k-1}{p-k}\cdot\frac{1}{C_{p-j}^{p-k}}} + 1 = \sum\limits_{j=1}^k \frac{1}{C_{p-(j+1)}^{p-k-1}} - \frac{p-k-1}{p-k}\cdot\sigma_p^{}(k) + 1.
\end{equation}
The first term can be transformed as follows:
\begin{equation}\nonumber
    \sum\limits_{j=1}^k \frac{1}{C_{p-(j+1)}^{p-k-1}} = \sum\limits_{j=2}^{k+1} \frac{1}{C_{p-j}^{p-k-1}} = \underbrace{\sum\limits_{j=1}^{k+1} \frac{1}{C_{p-j}^{p-k-1}}}_{\sigma_p^{}(k+1)} - \underbrace{\frac{1}{C_{p-1}^{p-k-1}}}_{1/C_{p-1}^k}.
\end{equation}
Thus, we have obtained the relation
\begin{equation}\nonumber
    \sigma_p^{}(k+1) = \sigma_p^{}(k+1) - \frac{1}{C_{p-1}^k} - \frac{p-k-1}{p-k} \cdot \sigma_p^{}(k) + 1,
\end{equation}
whence we find that
\begin{equation}\nonumber
    \sigma_p^{}(k) = \frac{p-k}{p-k-1} \left(1-\frac{1}{C_{p-1}^k}\right),
\end{equation}
and thereby 
\begin{equation}\nonumber
    \sum\limits_{j=1}^{m-1}\frac{1}{C_{m+n-j-1}^n} = \frac{n}{n-1} \left( 1 - \frac{1}{C_{m+n-2}^{m-1}} \right).
\end{equation}
Finally, taking into account Eqs.~\eqref{eq:B1}, \eqref{eq:B2} we have
\begin{equation}\nonumber
    \sum\limits_{j=1}^{m-1}\frac{C_{m+n-2}^{j-1}}{(m-j)\:C_{m-1}^{j-1}} = \frac{n}{n-1} \cdot \frac{C_{m+n-2}^{m-1}}{n} \cdot \left(1-\frac{1}{C_{m+n-2}^{m-1}}\right) = \frac{C_{m+n-2}^{m-1}}{n-1} - \frac{1}{n-1}.
\end{equation}
Then, instead of Eq.~\eqref{eq:A} for the coefficient $A$ we obtain
\begin{equation}\nonumber
    A = \frac{1}{2^m (n-1)}.
\end{equation}

\section{}
\label{sec:Appendix_C}

Let us show that
\begin{equation}\nonumber
    \sum\limits_{j=0}^m\sum\limits_{l=0}^n\frac{\left[(-1)^j+(-1)^l\right]C_m^j C_n^l}{j+l+1} = \frac{2^{m+n+1}}{m+n+1} \cdot \frac{1}{C_{m+n}^{n}} = \frac{2^{m+n+1}}{m+n+1} \cdot \frac{1}{C_{m+n}^{m}}
\end{equation}
for all $m,n\in\mathbb{N}\cup\{0\}$.

We transform the original double sum as follows:
\begin{equation}\nonumber
    \sum\limits_{j=0}^m\sum\limits_{l=0}^n\frac{\left[(-1)^j+(-1)^l\right]C_m^j C_n^l}{j+l+1} = \sum\limits_{j=0}^m\sum\limits_{l=0}^n\frac{(-1)^jC_m^j C_n^l}{j+l+1} + \sum\limits_{j=0}^m\sum\limits_{l=0}^n\frac{(-1)^lC_m^j C_n^l}{j+l+1} =
\end{equation}
\begin{equation}\label{eq:C_sum1}
    = \sum\limits_{l=0}^n C_n^l \sum\limits_{j=0}^m\frac{(-1)^jC_m^j}{j+l+1} + \sum\limits_{j=0}^m C_m^j \sum\limits_{l=0}^n\frac{(-1)^lC_n^l}{j+l+1}.
\end{equation}
Consider the first inner sum $\displaystyle\sum\limits_{j=0}^m\frac{(-1)^jC_m^j}{j+l+1}$. For fixed $m$, $l$ we introduce a polynomial
\begin{equation}\nonumber
    P(x) = P_{m,l}^{}(x) = \sum\limits_{j=0}^m\frac{(-1)^jC_m^j}{j+l+1}\:x^{j+l+1},
\end{equation}
for which $P(0)=0$ and the desired sum equals $P(1)$. Let's find the derivative
\begin{equation}\nonumber
    P^\prime(x) = \sum\limits_{j=0}^m (-1)^jC_m^j\:x^{j+l} = x^l \sum\limits_{j=0}^m (-1)^jC_m^j\:x^j = x^l \left( 1-x\right)^m.
\end{equation}
Then $P(x) = \displaystyle\int\limits_0^x t^l \left(1-t\right)^m dt$, and applying the Euler beta function, we get
\begin{equation}\label{eq:C_sum2}
    \displaystyle\sum\limits_{j=0}^m\frac{(-1)^jC_m^j}{j+l+1} = \displaystyle\int\limits_0^x t^l \left(1-t\right)^m dt =  \frac{l!\:m!}{(l+m+1)!}.
\end{equation}
Renaming $m$ to $n$ and swapping the roles of $j$ and $l$, from Eq.~\eqref{eq:C_sum2} we have
\begin{equation}\label{eq:C_sum3}
    \displaystyle\sum\limits_{l=0}^n\frac{(-1)^lC_n^l}{j+l+1} =   \frac{j!\:n!}{(j+n+1)!}.
\end{equation}
Returning to Eq.~\eqref{eq:C_sum1}, taking into account Eqs.~\eqref{eq:C_sum2} and \eqref{eq:C_sum3} we write
\begin{equation}\label{eq:C_sum4}
    \sum\limits_{j=0}^m\sum\limits_{l=0}^n\frac{\left[(-1)^j+(-1)^l\right]C_m^j C_n^l}{j+l+1} = \sum\limits_{l=0}^n C_n^l \frac{l!\:m!}{(l+m+1)!} + \sum\limits_{j=0}^m C_m^j \frac{j!\:n!}{(j+n+1)!}.
\end{equation}
Transforming the two sums separately, we get
\begin{equation}\nonumber
   \sum\limits_{l=0}^n C_n^l \frac{l!\:m!}{(l+m+1)!} = \frac{m!\:n!}{(m+n+1)!}\sum\limits_{l=0}^n\frac{(m+n+1)!}{(n-l)!\:(l+m+1)!} =
\end{equation}
\begin{equation}\nonumber
= \frac{1}{m+n+1}\cdot\frac{1}{C_{m+n}^n}\sum\limits_{l=0}^n C_{m+n+1}^{n-l} = \{k= m+1+l\} = \frac{1}{m+n+1}\cdot\frac{1}{C_{m+n}^n}\sum\limits_{k=m+1}^{m+n+1} C_{m+n+1}^{k},
\end{equation}
and similarly
\begin{equation}\nonumber
   \sum\limits_{j=0}^m C_m^j \frac{j!\:n!}{(j+n+1)!} = \frac{1}{m+n+1}\cdot\frac{1}{C_{m+n}^n}\sum\limits_{k=0}^{m} C_{m+n+1}^{k}.
\end{equation}
Hence the right-hand side of Eq.~\eqref{eq:C_sum4} takes the form
\begin{equation}\nonumber
   \frac{1}{m+n+1}\cdot\frac{1}{C_{m+n}^n}\left(\sum\limits_{k=0}^{m} C_{m+n+1}^{k}+\sum\limits_{k=m+1}^{m+n+1} C_{m+n+1}^{k}\right) = \frac{1}{m+n+1}\cdot\frac{1}{C_{m+n}^n}\sum\limits_{k=0}^{m+n+1} C_{m+n+1}^{k} =
\end{equation}
\begin{equation}\nonumber
    = \frac{2^{m+n+1}}{m+n+1}\cdot\frac{1}{C_{m+n}^n},
\end{equation}
and thus the required formula is obtained.

\section{}
\label{sec:Appendix_D}

The energy of a static kink can be found from Eq.~\eqref{eq:stenergy}. Taking into account that the kink solution satisfies Eq.~\eqref{eq:BPS}, and also using the expression for the potential \eqref{eq:potential_general}, we get
\begin{equation}\label{eq:D_1}
    M_{\scriptsize\mbox{K}}^{} = E[\varphi_{\scriptsize\mbox{K}}^{}(x)] = \int\limits_{-\infty}^{+\infty} 2V(\varphi_{\scriptsize\mbox{K}}^{}(x))\:dx = \int\limits_{-1}^{1} \sqrt{2V(\varphi)}\:d\varphi = \int\limits_{-1}^{1} \left(1+\varphi\right)^{m}\left(1-\varphi\right)^{n} d\varphi.
\end{equation}
A lot of combinatorial information about the double binomial in Eq.~\eqref{eq:D_1} is in Ref.~\cite{Petrosova.VMJ.2019}. By changing the variable $\displaystyle\frac{1+\varphi}{2}=t$, this integral reduces to the Euler beta function:
\begin{equation}\nonumber
    M_{\scriptsize\mbox{K}}^{} = 2^{m+n+1} \int\limits_0^1 t^m(1-t)^n dt = 2^{m+n+1} B\left(m+1, n+1\right).
\end{equation}
The final result can be written in one of the following three forms:
\begin{equation}\nonumber
    M_{\scriptsize\mbox{K}}^{} = \frac{2^{m+n+1}\:m!\:n!}{(m+n+1)!} = \frac{2^{m+n+1}}{m+n+1} \cdot \frac{1}{C_{m+n}^{n}} = \frac{2^{m+n+1}}{m+n+1} \cdot \frac{1}{C_{m+n}^{m}},
\end{equation}
which, of course, is the same as Eq.~\eqref{eq:mass_general}. In the special case $m=1$ this gives the result \eqref{eq:mass_for_m_equals_1}, and for $m=n$ we get Eq.~\eqref{eq:mass_for_m_equals_n}.

\end{document}